\documentclass[11pt,a4paper,english]{article}
\usepackage[T1]{fontenc}
\usepackage[utf8]{inputenc}
\usepackage{babel}

\usepackage[sorting=none]{biblatex}
\usepackage{caption}
\usepackage{blindtext}
\usepackage{amssymb}
\usepackage[export]{adjustbox}
\usepackage{amsmath}
\usepackage{graphicx}
\usepackage{float}
\usepackage{appendix}
\usepackage{lipsum}
\usepackage{hyperref}
\usepackage{csquotes}
\hypersetup{
    colorlinks=true,
    linkcolor=blue,
    filecolor=blue,      
    urlcolor=blue,
    citecolor=blue,
}

\title{Deep Learning for Digital Asset Limit Order Books}
\author{%
    Rakshit Jha \\
    \small \href{https://www.cam.ac.uk}{University of Cambridge}\\
    \and
    Mattijs De Paepe \\
    \small \href{https://www.cam.ac.uk}{University of Cambridge}\\
    \and
    Samuel Holt \\
    \small \href{mailto:samuel.holt.direct@gmail.com}{samuel.holt.direct@gmail.com}
    \and
    James West\\
    \small \href{https://globedx.com}{Globe Research}\\
    \and
    Shaun Ng\\
    \small \href{https://globedx.com}{Globe Research}
}

\begin{document}
\maketitle

\begin{abstract} \noindent 
This paper shows that temporal CNNs accurately predict bitcoin spot price movements from limit order book data. On a 2 second prediction time horizon we achieve 71\% walk-forward accuracy on the popular cryptocurrency exchange coinbase. Our model can be trained in less than a day on commodity GPUs which could be installed into colocation centers allowing for model sync with existing faster orderbook prediction models. We provide source code and data at \url{https://github.com/Globe-Research/deep-orderbook}.
\end{abstract}

\section{Introduction}
Digital assets (cryptocurrencies) have a more relaxed regulatory regime than most traditional financial instruments. Due to the nasent technologies typically used, cryptocurrency exchanges also operate at much higher latencies than traditional assets, tail exchange response time latencies often spanning into several seconds, rather than microseconds. As a consequence, their orderbooks are likely to contain complex phenomena that are less common in traditional markets. Due to the longer latencies and richer orderbook phenomena, deep learning becomes a much more viable tool in predicting cryptocurrency limit order book (LOB) dynamics \cite{zhang_zohren_roberts_2019}.

Prior work using machine learning techniques in equities and futures markets, suggests that they are likely to be effective predicting cryptocurrency price movements. Indeed, many papers showed this to be true by using Bayesian neural networks \cite{jang_lee_2017}, gradient boosting decision trees \cite{alessandretti_elbahrawy_aiello_baronchelli_2018}, long short-term memory neural networks \cite{mcnally_roche_caton_2018}, and many other algorithms. Aside from improved predictive performance, machine learning techniques also allow a probabilistic interpretation which classical time series forecasting methods such as ARIMA models lack. With the high volatility of cryptocurrencies, risk management is even more important. As with many fields now, deep learning techniques are increasingly yielding state of the art performance.

Many studies using LOB extract relevant features from the LOB as it is otherwise too high-dimensional for a 'regular' implementation of a neural net as they do not scale well to high-dimensional inputs. The most common solution to this problem, first developed for analysing visual imagery and inspired by our visual cortex, is the Convolutional Neural Network (CNN). Prior work comparing a CNN approach with a traditional Multi-Layer Perceptron (MLP) for traditional stocks showed it led to better results for many prediction horizons \cite{tsantekidis_passalis_tefas_kanniainen_gabbouj_iosifidis_2017} (see also \cite{zhang_zohren_roberts_2019} and \cite{hernandez-2018}).

Sirignano\cite{sirignano2016deep} introduced spatial neural networks as a low-dimensional means of predicting the limit order book at a future time conditional on the current state of the limit order book, incorporating market information deep in the book in a computationally feasible manner (50 GPU clusters at the time).

This paper illustrates the viability of CNN-like architectures on the kind of commodity hardware that a high frequency trader could install into colocation, to show that cryptocurrencies could benefit from such an approach.

\section{Methodology}
\subsection{Orderbook and midprice data}

We use a high frequency limit order book (see, e.g. \cite{paraskevi_2019}) data at 100ms precision, due to the exchange latency. Limit order book data is from the cryptocurrency exchange coinbase, an established spot trading platform for bitcoin to dollar exchange rates. The data spans 9 days with 50 levels of non-zero best bid and best ask. The minimum price increment on coinbase is much smaller than traditional futures and equities exchanges, resulting in different dynamics, fewer market participants typically offering quotes on a given price level, and most price levels have zero depth away from market midprice.

\begin{figure}[H]
\begin{tabular}{ccc}
    \includegraphics[width=0.3\textwidth]{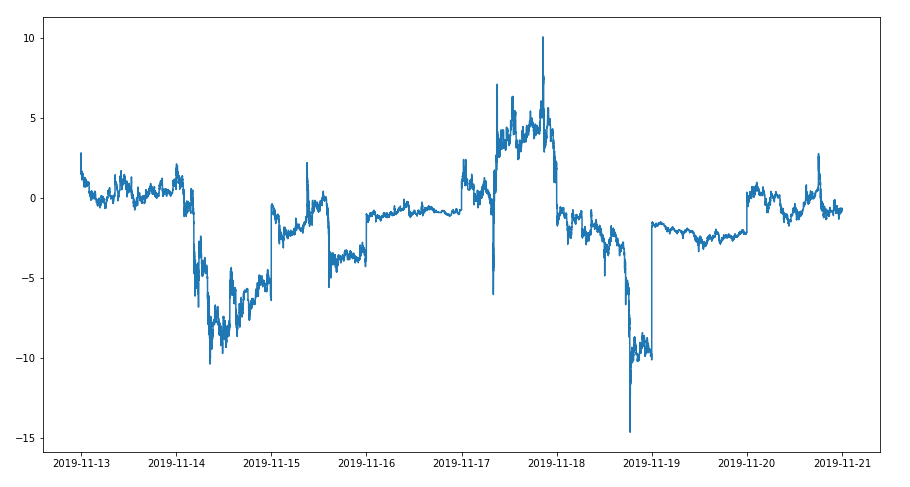} & \includegraphics[width=0.3\textwidth]{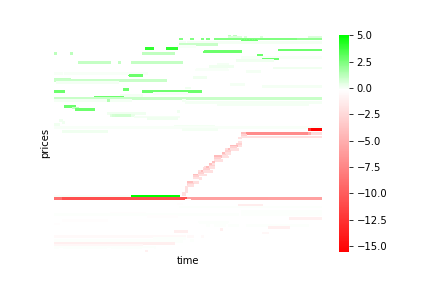} & \includegraphics[width=0.3\textwidth]{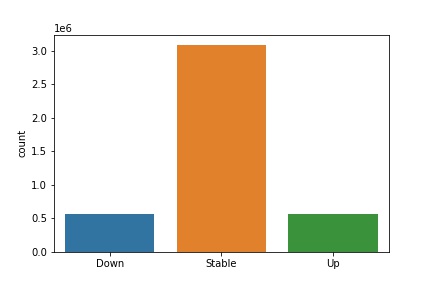} \\
    (a) BTCUSD midprice & (b) Sample feature data (c) & Class balance pre-downsampling \\
\end{tabular}
\caption{
    Summary of 100ms resolution price and orderbook data from coinbase
    (a) Sample BTCUSD exchange rate orderbook data from coinbase: 10 seconds total span covering multiple levels of bids and asks
    (b) Example orderbook features used to train model: asks in red, bids in green
    (c) Class imbalance resolution
}
\label{fig:input}
\end{figure}

The data contains snapshots every 100ms up to a depth of 50 for 9 consecutive days, from the 12th to the 20th of June 2019. Each order is described by its price and volume. A depth of 50 for both bid and ask thus means each timestep contains 200 features. The data is near continuous and not discretized ticks. Figure \ref{fig:input} shows the midprice evolution during the time period. 

The range of the price and volume is wildly different and not of a suitable magnitude for our neural network so we first normalise the data. We standardize the data by using the mean and standard deviation of the previous day's prices and volumes i.e. $ \textbf{x}_{norm} = \frac{\textbf{x} - \bar{x}}{\sigma_x}$. This is because the distributions of both price and volume can shift significantly in a few days time.

While Sirignano \cite{sirignano2016deep} found phenomena deep in the orderbook for equities (around 25 levels) and others have found phenomena around 10 levels deep for futures \cite{tsantekidis_passalis_tefas_kanniainen_gabbouj_iosifidis_2017}, we found the majority of value within the top 10 bid and ask price levels as the very minor performance improvement was not worth the increased computational cost, see figure \ref{fig:walkforwadparams} (a).

Existing literature has often taken 40 features and 100 timesteps e.g. in \cite{tsantekidis_passalis_tefas_kanniainen_gabbouj_iosifidis_2017} and \cite{wallbridge20}.
In figure \ref{fig:input}, we see solid green lines at top of the prices as buying is usually set higher and for asks we see the opposites as selling is cheaper. Such instances are fed into the model to predict mid price changes. The colour bar shows negative values as that was used to distinguish between buy and sell and the data is not normalised while creating this heat map. 

We look to predict movement in the midprice of the bitcoin to dollar exchange rate, defined as usual as $p_t = {(p_a(t) + p_b(t))}/{2}$ where $p_a(t)$ is the price of the best ask at time $t$ and similarly $p_b(t)$ is the price of the best bid at $t$. Defining the $k$-averaged midprices $m_{-/+}$ before and including time $t$ ($-$) and immediately after time $t$ ($+$) as 

\begin{equation}
    m_{-}(t) = \frac{1}{k}\sum_{i=0}^{k-1} p_{t-i} 
\end{equation}
\begin{equation}
    m_{+}(t) = \frac{1}{k}\sum_{i=1}^{k} p_{t+i}
\end{equation}

We classify according to the signed movements in average midprice between $\{t--k(-1), ..., t-1, t\}$ and $\{t+1, t+2, ... t+k\}$. In our case $k=20$.

\begin{equation}
    l_t = \begin{cases}
    1 &\text{if $m_-(t)>m_+(t)(1+\alpha)$}\\
    -1 &\text{if $m_-(t)<m_+(t)(1-\alpha)$}\\
    0 & \text{otherwise}\\
    \end{cases}
\end{equation}

We took $\alpha = 0.002$ after an exhaustive search, consistent with similar studies e.g. \cite{wallbridge20}. Since midprices rarely change every 100ms, the resulting $l_t$ were highly class imbalanced as shown in Figure \ref{fig:input}. To resolve this we randomly downsampled to remove excessive price-constant (Stable) events i.e. $l_0$ from the data. The total samples decreased from $4,219,932$ before balancing to $1,681,407$ after downsampling.

\subsection{Model Architecture}
We use a Temporal Convolution Neural Networks \cite{BaiTCN2018} for time series prediction. TCNs have been designed from two basic principles, which are very crucial to the time series predictions:
\begin{itemize}
    \item The padding for convolutions are causal. This prevents any information leakage from future to past. 
    \item The architecture can take a sequence of any length and map it to an output sequence of the same length just as with an RNN. 
\end{itemize}

We chose TCN over their counterparts like GRU and LSTM for reasons such as low memory requirement, parallelism, flexibility to play around with receptive fields gave us a huge reduction in training time, considering the dataset size at 100ms resolution. 

We used a kernel size of 2 and calculated a dilation number of 6 from the relation
\begin{equation*}
    \mathrm{timesteps} = 1 + 2({\mathrm{kernel\_size}-1)(2^{\mathrm{dilation\_number}}-1)}
\end{equation*}

\footnotetext{See TF2.0 Implementation of TCN package on \cite{philipperemy}}

\subsection{Evaluation}
\subsubsection{Training}
\label{subsec:training}
The parameters of the model are learnt by minimising the categorical cross entropy loss. ADAM is used as an optimizer with setting the learning rate as 0.01 and $\epsilon$ as $10\text{e-}7$. The early stopping callback is used to stop training when the validation loss doesn't improve after 4 epochs. We also use ReduceLROnPlateau\cite{chollet2015keras}, which reduces the learning rate by a factor of 0.5 if the validation loss doesn't improve for 2 epochs. We train with a batch size of 128. 

\subsubsection{Validation}
The model is of an appropriate size as demonstrated by figure \ref{fig:cross_entropy}. Any over- or under-fitting would show up on the loss curves and would indicate a potential need of a strategy to prevent overfitting such as regularization or dropout. After training for 57 epochs using the method described in section \ref{subsec:training}, we present our results as a confusion matrix of the recall on the test dataset, as can be seen in figure \ref{fig:confusion_matrix}.

\section{Results}
We use a walkforward approach as depicted in figure \ref{fig:walkforward_summary}, testing out training periods of length 1 day to 7 days, but keeping the test period as 1 day and walking forwards through the dataset. 

\subsection{Strong predictive accuracy}

Unsurprisingly, it was far easier to predict price stability - even in a downsampled dataset with class balance - but with sufficient predictive accuracy to aid a market making algorithm, yielding around 61\% accuracy on downward movements on a  2 second horizon, 66\% for upward price movements using a 7 day training, 1 day testing (after training period) split. Overall performance of the temporal CNNs in cryptocurrencies here is comparable to the results of Siginano \cite{sirignano2016deep} in equities.

\begin{figure}[H]
\begin{tabular}{cc}
    \raisebox{.5\height}{\includegraphics[width=0.45\textwidth]{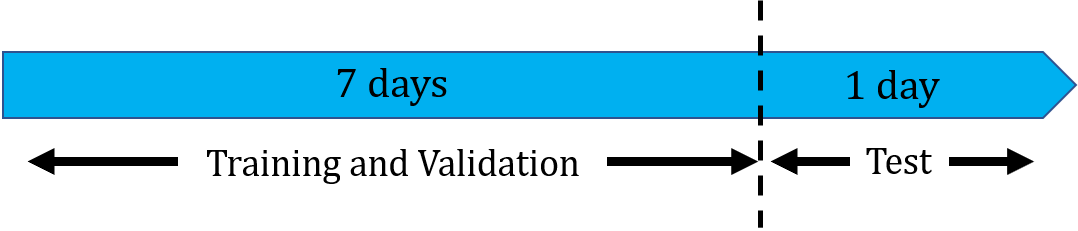}}
    & \includegraphics[width=0.45\textwidth]{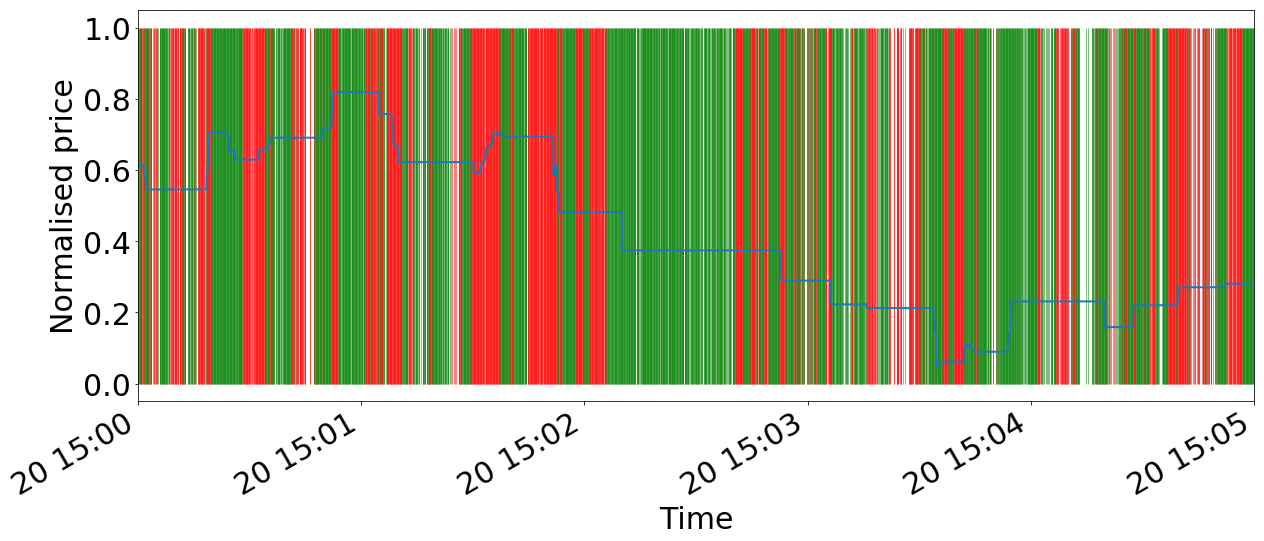} \\
    (a) Study design & (b) Classification correctness \\
    \raisebox{-.5\height}{\includegraphics[width=0.35\textwidth]{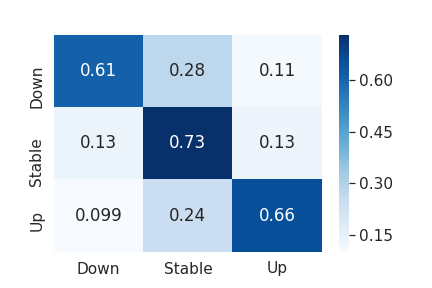}} & 
            \resizebox{0.5\textwidth}{!}{
                \begin{tabular}{|c|c c c|c|c|}
                     \hline
                     \textbf{Class} &  \textbf{Precision} &  \textbf{Recall} &  \textbf{F1} & \textbf{Accuracy} & \textbf{Support}\\
                     \hline
                     Down   & 0.42 & 0.61 & 0.50 & 61\% & 61142  \\
                     Stable & 0.90 & 0.73 & 0.81 & 73\% & 350499 \\
                     Up     & 0.40 & 0.66 & 0.50 & 66\% & 53759  \\
                     \hline
                     Avg. & 0.78 & 0.71 & 0.73 & 71\% & 465400\\
                     \hline
                \end{tabular}
            }
    \\
    (c) Confusion matrix & (d) Forward performance \\
\end{tabular}
\caption{
        Walkforward study design, classification outcomes and forwards performance stability
        a. Study design 
        b. Classification summary 
        c. Confusion matrix 
        d. Forwards performance stability (green = correct, red = wrong prediction)
}
\label{fig:walkforward_summary}
\end{figure}

Whilst our dataset is limited, we find in general that models do not "wear off" noticeably in accuracy over the course of a day, suggesting much slower market response to existing predictability (Figure \ref{fig:walkforward_summary} (b)). This points to far less regular model refitting being required, alongside the improved model performance seen with longer training windows (Figure \ref{fig:walkforwadparams}

\subsection{Market and model mechanics}
\begin{figure}[H]
\begin{tabular}{ccc}
    \includegraphics[width=0.33\textwidth]{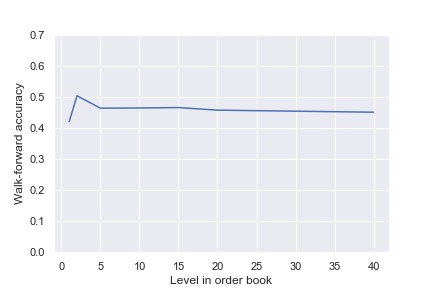} & \includegraphics[width=0.33\textwidth]{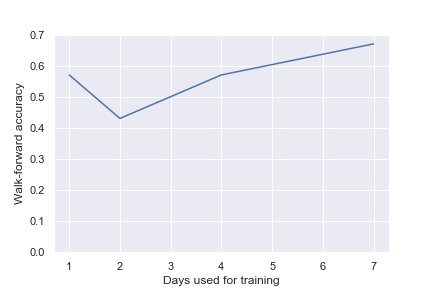} & \includegraphics[width=0.33\textwidth]{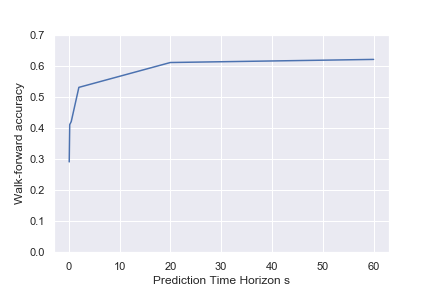} \\
    (a) Order book Level depth & (b) Days used for training & (c) Prediction horizon (seconds)   \\
\end{tabular}
\caption{
        Walkforward accuracy study varying the hyper parameters
        a. Order book Level depth
        b. Days used for training
        c. Prediction time horizon (seconds)
}
\label{fig:walkforwadparams}
\end{figure}

We explore how varying the predictive time horizon, depth of book  and training time period, using the model as a lens on the mechanics of the market in Figure \ref{fig:walkforwadparams}. Notably:
\begin{enumerate}
    \item Orderbook phenomena on coinbase appear quite shallow in the book, without any performance gains past 10 levels, and an overall decline due to what should best be interpreted as "noise features" from the model's perspective.
    \item In contrast to classic logistic regression models of price movements based on orderbook data for futures and equities on e.g. the CME or NASDAQ, model performance improves significantly with a long training period, compared to the intraday retraining often used for other asset classes.
    \item Predictability of midprice movement remains for quite a while, often for up to a minute on coinbase.
\end{enumerate}

Overall, we find rather different behaviour in cryptocurrency orderbooks compared to traditional assets. Given the architecture of the model, it is likely capturing the persistence and relative strength of orderbook imbalance for the 5 or so best bids and asks, likely offering better performance than the appropriate logistic regression model equivalent, although also likely requiring a longer training period (still increasing at the limit of the length of our dataset).

\section{Conclusion}
Deep learning techniques have a long history in quantitative trading in traditional finance, but their real world applications to high frequency trading are relatively new. In the spirit of Tinbergen, we present a short list of stylized facts about their application:
\begin{enumerate}
    \item Deep learning models provide higher latency predictions than traditional machine learning techniques in predicting price movements from historic data.
    \item Whilst we don't explicitly validate that this is the case for cryptocurrency orderbooks, it is well established for traditional assets that deep learning models provide more accurate forecasts of price action from historic orderbook movements.
    \item Deep learning models are, as of 2020, viable for use in colocation centers - trainable within a sufficiently short time horizon on affordable commodity GPUs to catch the rapid market regime evolution typical of financial instruments (hours to days).
    \item Market regime durations for cryptocurrency orderbook phenomena are hours to days.
    \item Cryptocurrency orderbooks have a "memory" of at least the last 10 seconds in existing spot markets.
\end{enumerate}

The temporal CNNs used in the paper are relatively nascent, and we expect further improvements on their architecture, training etc. in the future which will only further benefit market participants looking to deploy such technology. 

We believe that further extensions of this work are likely to look at 
\begin{enumerate}
    \item The joint distributions between the orderbooks of the emerging derivatives markets for digital assets and their spot-market equivalent orderbooks.
    \item The regime durations of these markets and the likely improvements of move online modelling e.g. via online training downweighting older events in stochastically sampled batches.
    \item The orderbook depth at which price-movement predictive phenomena occur for cryptocurrency spot markets is about 10 levels, comparable to futures markets, and shallower than equities markets, although this may change.
    
\end{enumerate}

\section{Acknowledgements}

The research in this paper was made possible by resources provided by \href{https://globedx.com}{Globe Research} as part of Globe, a pioneering cryptocurrency derivatives exchange, available at \url{https://globedx.com}.

\newpage

\nocite{*}
\printbibliography

@article{tsantekidis_passalis_tefas_kanniainen_gabbouj_iosifidis_2017,
    title={Forecasting Stock Prices from the Limit Order Book Using Convolutional Neural Networks},
    DOI={10.1109/cbi.2017.23},
    journal={2017 IEEE 19th Conference on Business Informatics (CBI)},
    author={Tsantekidis, Avraam and Passalis, Nikolaos and Tefas, Anastasios and Kanniainen, Juho and Gabbouj, Moncef and Iosifidis, Alexandros},
    year={2017}}

@article{sirignano2016deep,
    title={Deep Learning for Limit Order Books},
    author={Justin Sirignano},
    year={2016},
    eprint={1601.01987},
    archivePrefix={arXiv},
    primaryClass={q-fin.TR}
}

@article{BaiTCN2018,
	author    = {Shaojie Bai and J. Zico Kolter and Vladlen Koltun},
	title     = {An Empirical Evaluation of Generic Convolutional and Recurrent Networks for Sequence Modeling},
	journal   = {arXiv:1803.01271},
	year      = {2018},
}

@article{jang_lee_2017,
    title={An Empirical Study on Modeling and Prediction of Bitcoin Prices With Bayesian Neural Networks Based on Blockchain Information},
    volume={6},
    DOI={10.1109/access.2017.2779181},
    journal={IEEE Access},
    author={Jang, Huisu and Lee, Jaewook},
    year={2017},
    month={12},
    pages={5427–5437}}

@article{paraskevi_2019,title={Machine Learning for Forecasting Mid Price Movement using Limit Order Book Data},author={Nousi, Paraskev and Tsantekidis, Avraam and Passalis, Nikolaos and Tefas, Anastasios and Kanniainen, Juho and Gabbouj, Moncef and Iosifidis, Alexandros},year={2019},month={Apr}}

@article{mcnally_roche_caton_2018,
    title={Predicting the Price of Bitcoin Using Machine Learning},
    DOI={10.1109/pdp2018.2018.00060},
    journal={2018 26th Euromicro International Conference on Parallel, Distributed and Network-based Processing (PDP)},
    author={Mcnally, Sean and Roche, Jason and Caton, Simon},
    year={2018}}

@article{alessandretti_elbahrawy_aiello_baronchelli_2018,
    title={Anticipating Cryptocurrency Prices Using Machine Learning},
    volume={2018},
    DOI={10.1155/2018/8983590},
    journal={Complexity},
    author={Alessandretti, Laura and Elbahrawy, Abeer and Aiello, Luca Maria and Baronchelli, Andrea},
    year={2018},
    month={04},
    pages={1–16}}

@article{he_zhang_ren_sun_2016,
    title={Deep Residual Learning for Image Recognition},
    DOI={10.1109/cvpr.2016.90},
    journal={2016 IEEE Conference on Computer Vision and Pattern Recognition (CVPR)},
    author={He, Kaiming and Zhang, Xiangyu and Ren, Shaoqing and Sun, Jian},
    year={2016}}

@article{zhang_zohren_roberts_2019,
    title={DeepLOB: Deep Convolutional Neural Networks for Limit Order Books},
    volume={67},
    DOI={10.1109/tsp.2019.2907260},
    number={11},
    journal={IEEE Transactions on Signal Processing}, author={Zhang, Zihao and Zohren, Stefan and Roberts, Stephen},
    year={2019},
    month={01},
    pages={3001–3012}}

@article{hernandez-2018,
    title={CNN with Limit Order Book Data for Stock Price Prediction},
    DOI={10.1007/978-3-030-02686-8_34},
    journal={Proceedings of the Future Technologies Conference (FTC) 2018 Advances in Intelligent Systems and Computing},
    author={Niño, Jaime and Hernandez, German and Arévalo, Andrés and Leon, Diego and Sandoval, Javier},
    year={2018},
    pages={444–457}}

@article{wallbridge20,
    author  = {J. Wallbridge},
    title   = {Transformers for Limit Order Books},
    year    = {2020},
    journal = {arXiv:2003.00130},
    volume  = {},
    pages   = {},
    eprint  = {arXiv:2003.00130}
}

@misc{philipperemy,
title={Keras-TCN},
author={Philippe Rémy},
year={2020},
howpublished={\url{https://github.com/philipperemy/keras-tcn}},
}

@misc{chollet2015keras,
  title={Keras},
  author={Chollet, Fran\c{c}ois and others},
  year={2015},
  howpublished={\url{https://keras.io}},
}

@misc{tensorflow2015-whitepaper,
    title={ {TensorFlow}: Large-Scale Machine Learning on Heterogeneous Systems},
    url={https://www.tensorflow.org/},
    note={Software available from tensorflow.org},
    author={
    Martin~Abadi and
    Ashish~Agarwal and
    Paul~Barham and
    Eugene~Brevdo and
    Zhifeng~Chen and
    Craig~Citro and
    Greg~S.~Corrado and
    Andy~Davis and
    Jeffrey~Dean and
    Matthieu~Devin and
    Sanjay~Ghemawat and
    Ian~Goodfellow and
    Andrew~Harp and
    Geoffrey~Irving and
    Michael~Isard and
    Yangqing Jia and
    Rafal~Jozefowicz and
    Lukasz~Kaiser and
    Manjunath~Kudlur and
    Josh~Levenberg and
    Dandelion~Man\'{e} and
    Rajat~Monga and
    Sherry~Moore and
    Derek~Murray and
    Chris~Olah and
    Mike~Schuster and
    Jonathon~Shlens and
    Benoit~Steiner and
    Ilya~Sutskever and
    Kunal~Talwar and
    Paul~Tucker and
    Vincent~Vanhoucke and
    Vijay~Vasudevan and
    Fernanda~Vi\'{e}gas and
    Oriol~Vinyals and
    Pete~Warden and
    Martin~Wattenberg and
    Martin~Wicke and
    Yuan~Yu and
    Xiaoqiang~Zheng},
    year={2015},
}

@article{ntakaris_magris_kanniainen_gabbouj_iosifidis_2018,
    title={Benchmark dataset for mid-price forecasting of limit order book data with machine learning methods},
    volume={37},
    DOI={10.1002/for.2543},
    number={8},
    journal={Journal of Forecasting},
    author={Ntakaris, Adamantios and Magris, Martin and Kanniainen, Juho and Gabbouj, Moncef and Iosifidis, Alexandros},
    year={2018},
    pages={852–866}}

\appendix
\section{Model Architecture}
\begin{figure}[H]
    \centering
    \includegraphics[height=0.90\textheight]{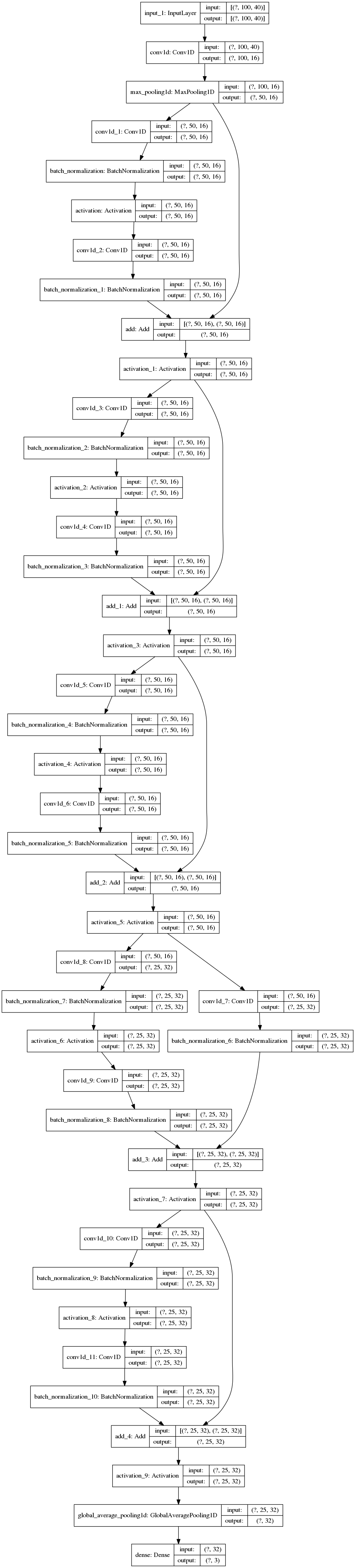}
    \caption{Model in the paper}
    \label{fig:model_architecture_resnet}
\end{figure}

\section{Model training}
The confusion matrix for the predictions on the test set can be seen in figure \ref{fig:confusion_matrix} and the classification report in table \ref{tab:classification_report}.

\begin{minipage}[c]{.4\textwidth}
\begin{figure}[H]
    \centering
    \includegraphics[width=\textwidth]{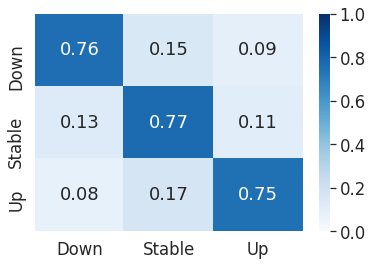}
    \caption{Confusion Matrix}
    \label{fig:confusion_matrix}
\end{figure}
\end{minipage}
\begin{minipage}[c]{.6\textwidth}
\begin{table}[H]
    \resizebox{\textwidth}{!}{
    \begin{tabular}{|c|c c c|c|}
         \hline
         \textbf{Classes} &  \textbf{Precision} &  \textbf{Recall} &  \textbf{F1-Score} & \textbf{Accuracy} \\
         \hline
         Down (0) & 0.78 & 0.76 & 0.77  & \\
         Stable (1) & 0.71 & 0.77 & 0.74 & \\
         Up (2) & 0.79 & 0.75 & 0.77 & \\
         \hline
         Macro Avg. & 0.76 & 0.76 & 0.76 & 0.76 \\
         Weighted Avg. & 0.76 & 0.76 & 0.76  & \\
         \hline
    \end{tabular}}
    \caption{Classification Report}
    \label{tab:classification_report}
\end{table}
\end{minipage}

\begin{minipage}{.5\textwidth}
\begin{figure}[H]
    \centering
    \includegraphics[width=\textwidth]{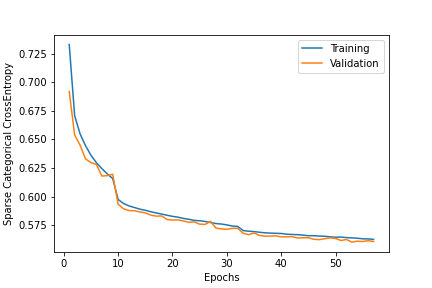} 
    \caption{Cross Entropy Loss Plot}
    \label{fig:cross_entropy}
\end{figure}
\end{minipage}
\begin{minipage}{.5\textwidth}
\begin{figure}[H]
    \centering
    \includegraphics[width=\textwidth]{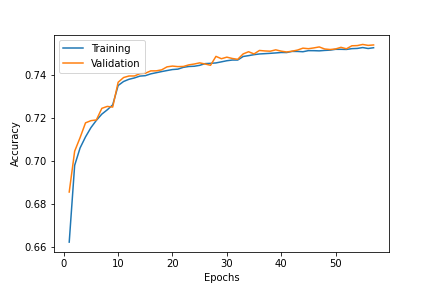}
    \caption{Accuracy Plot}
    \label{fig:accuracy}
\end{figure}
\end{minipage}

\end{document}